# Uneven Current Distribution Modeling and Random Walk based Variation Analysis for interconnect Electromigration Estimation

Karthik Airani, Rohit Guttal
Department of Silicon Engineering, Micron Technology
8000 South Federal Way, Boise, Idaho

**Abstract**
In this paper, we modeled and simulated uneven current distribution for on-die interconnect structure. We show significant difference when considering uneven current distribution. Finite Element Method approach is used to analyze various variation effects. Blech length effect and analytical EM lifetime calculation is used to quickly identify on-die EM weak wires. We propose a random walk based approach to evaluate on-die variation impact on electromigration reliability. Experimental results show substantial EM degradation with variation effects present.

## 1. Introduction

With technology scaling, on-die current density keeps increasing, thus posts huge challenge on chip electromigration closure. Moreover, variation effects such as CMP dishing and lithography edge placement error are becoming more and more important with advanced technology node. All those effects impact chip electromigration reliability significantly. The EM degradation needs to be captured during power grid integrity analysis to guarantee overall chip reliability. In this paper we analyze uneven current distribution of on-die interconnect structure and use random walk to simulate variation effects impact.

## 2. Current Distribution Model
### 2.1 Interconnect Structure

The interconnect structure that affects EM mainly refers to the via-array design. For a single via, all current is flowing through the via and its current density can be easily obtained. However in a via-array, the current is not evenly distributed in- to each via. References [1]-[2] discuss this effect.

In [3], a 3D orthogonal wire structure with a 4-by-4 via-array is built as shown in Figure 1 (a). Several current configurations are applied to the structure, and current densities through the horizontal cross-sectional plane of multiple vias are recorded as shown in Figure 1 (b)-(d). It can be observed that when all wire cur- rents are identical as in Figure 1 (b), current distribution into the vias is also uni- form. Greater differences among the wire segment currents cause the current split into the vias to be more non-uniform.

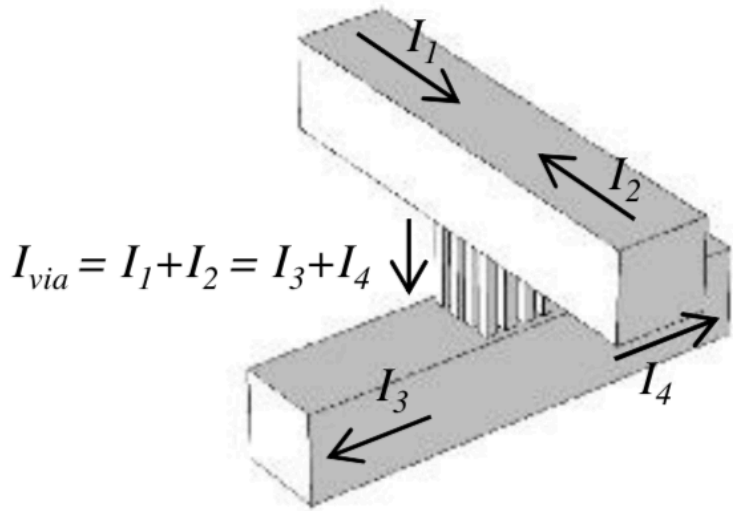

(a) 3D orthogonal wire structure with 4-by-4 via-array.

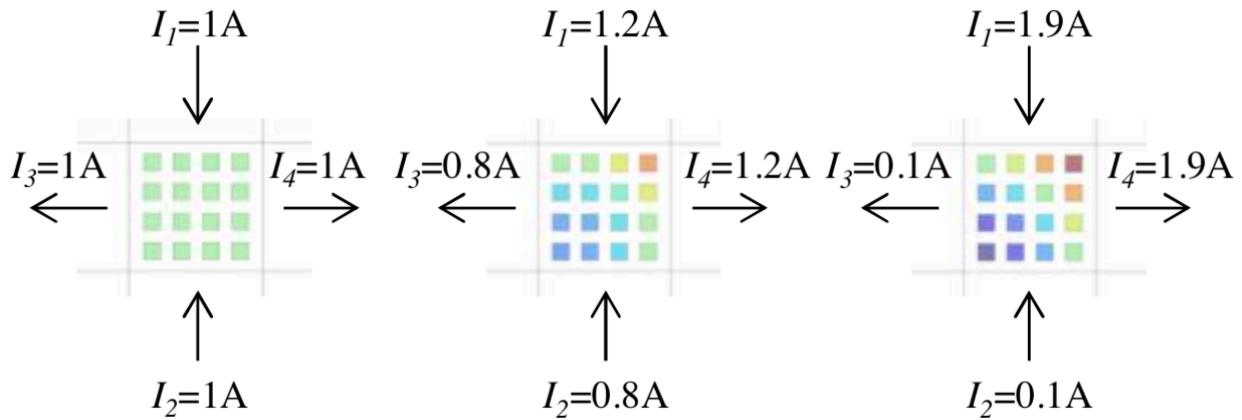

(b) Even distribution  (c) Uneven distribution  (d) Extreme uneven distribution.

Fig. 1. Via-array current distribution.

This effect can be explained by the differences among path resistances through different vias. A SPICE simulation on a pure resistive mesh validates this observation. Based on the resistive mesh model, the current distribution into each via can be determined and the EM lifetime of the via-array can be estimated. A detailed discussion can be found in [4].

## 2.2 Activation Energy

Activation energy is the minimum energy required to cause thermal diffusion of metal ions. It plays an important role due to its exponential effect on MTTF [5]. Figure 2 shows the MTTF trend when activation energy changes, while temperature is fixed at 300K. In Figure 2, MTTF is normalized to its value at $E_a=0.7eV$. The exponential dependency can be observed.

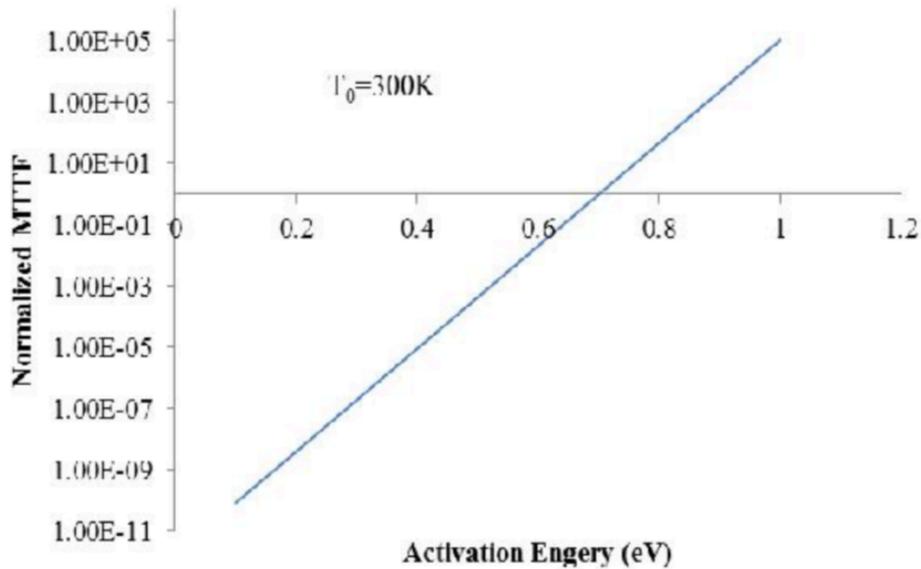

**Fig. 2. MTTF-Activation energy dependency**

Interconnect material is the dominant factor that determines the value of $E_a$. Aluminum was the typical material used for interconnects before copper was first introduced in 1997 [6]. Due to its generally higher activation energy, copper has a significantly better electromigration resistance than aluminum does. In this chap- ter, we focus our discussion on copper interconnect EM reliability.

# 3. Physical Simulatin of EM
## 3.1 Balance of Atom Concentration

The governing equation which describes the atom concentration evolution throughout an interconnect segment, is the conventional mass balance (continuity) equation in (1) [6].

$$\frac{\partial N(x, t)}{\partial t} + \nabla \cdot J = 0 \tag{1}$$

In (1), N(x, t) is the atom concentration at the point with coordinates x=(x, y, z) at the moment of time t, and J is the total atomic flux at this location. The total atomic flux J is a combination of the fluxes caused by the different atom migration forces. The major forces are induced by the electric current, and by the gradients of temperature, mechanical stress and concentration: J =$J_E$+$J_{th}$+$J_S$+$J_C$.

To define the fluxes mentioned above, we have (2).

$$\begin{cases} \vec{J}_E = \frac{N}{kT} eZ^* j\rho D_0 e^{-\frac{E_a}{kT}} \\ \vec{J}_{th} = -\frac{NQ^* D_0}{kT^2} e^{-\frac{E_a}{kT}} gradT \\ \vec{J}_s = -\frac{N\Omega D_0}{kT} e^{-\frac{E_a}{kT}} grad\sigma_H \\ \vec{J}_c = -D_0 e^{-\frac{E_a}{kT}} gradN \\ \frac{dN}{dt} + div(\vec{J}_E + \vec{J}_{th} + \vec{J}_s + \vec{J}_c) = 0 \end{cases} \tag{2}$$

In (2), $J_E$, $J_{th}$, $J_S$ and $J_C$ are atomic fluxes induced by electrical, thermal, stress and atomic concentration forces, respectively. N is the atomic concentration, $Q^*$ is the specific heat of transport of metal, and $\sigma_H$ is the local hydrostatic stress. EM lifetime is considered to be proportional to atomic flux divergence (AFD=divJ), so we have here MTTF∝AFD.

## 3.2 Simulation Setup and Result

The 3D interconnect structure used for simulation is a copper wire segment with a via below the wire. The wire geometry is specified by its 5μm width, 100μm length and 1.5μm height; the via is of 2.5μm size and is 1.5μm off the wire edge; the initial temperature is set to 300K; the activation energy is set to 0.7eV for the copper body and is set to 1.25eV along the surface to mimic the behavior of the interface between the copper body and the cap layer. The current is flowing from the via bottom up to the wire, then to the other end of the wire. A 3D view is shown in Figure 3.

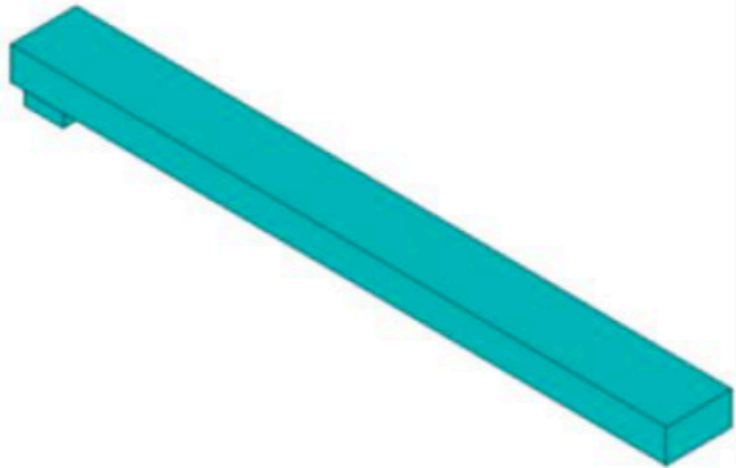

**Fig. 3. 3D view of a simulated wire structure.**

We build an EM physical simulator using ANSYS [7] and a C++ program. ANSYS is a multi-physics FEM-based tool, which provides AFD distribution for specified physical parameters including temperature, current density and stress. The whole simulation time is divided into many small time steps $\Delta t$. During each time step, ANSYS simulates the wire structure and C++ program calculates the AFD using (2), then the AFD is fed back into ANSYS for simulation of the next time step. This ANSYS/C++ loop continues until an EM failure state is reached (e.g. 10% resistance increase).

Figure 4 shows the AFD contour map of the interconnect structure. The wire end that connects to the via below has large AFD, while small AFD is found at the other end. The result matches well with the theoretical analysis and experimental observations from real EM tests in literature [8-9].

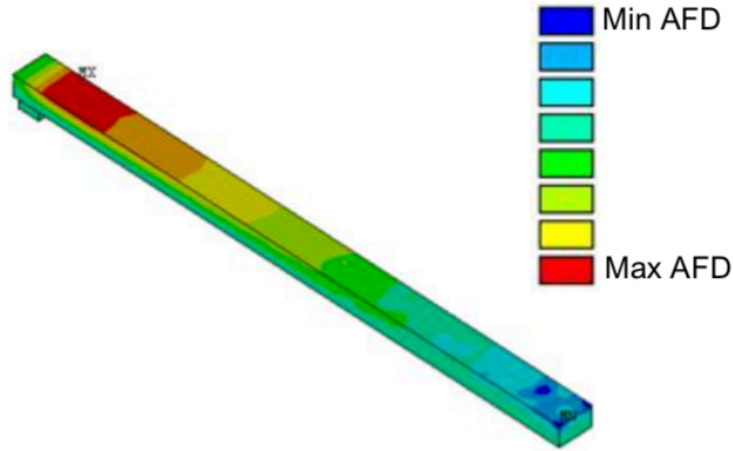

**Fig. 4. Physical simulation result of a wire segment**

## 4. Power Grid EM Evaluation
### 4.1 Power Grid Model
Power grid of a VLSI chip carries unidirectional currents of large magnitudes and it is the most EM-vulnerable part of the on-chip interconnect network. In this chapter, we focus on power grid wires only.

In Figure 5, we show the circuit-level model of power grid used in our work. We assume that power is delivered through a controlled collapse chip connection (C4) package. Wires and vias are modeled as resistors, external voltage supplies are modeled as ideal voltage sources, and switching transistors are modeled as current sources.

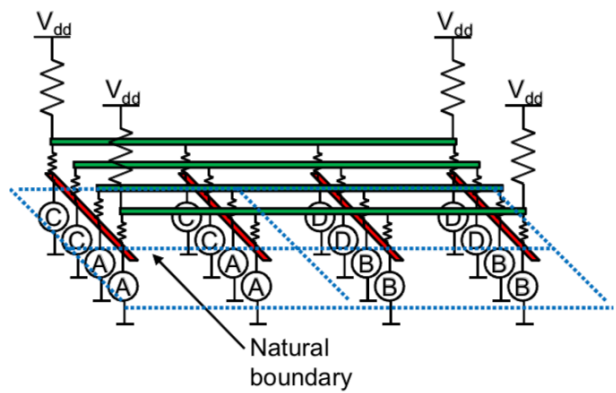

**Fig. 5 Circuit-level power grid model.**

There are many different power grid EM analysis flows depending on data format types used to describe the power grid. In this chapter, we use the SPICE format. The link between the SPICE netlist naming and numbering scheme for the circuit and the original geometry of the power grid is described below.

· Node name:
n<net-index>_<x-location>_<y-location>
· Data associated with each layer starts from:
* layer: <name>,<net>_net: <net-index>
Each layer/net combination is associated with a unique net-index. · Vias starts from:
* vias from: <net-index> to <net-index>

Vias are implemented as resistors or as zero voltage sources.
· Current source:
iB<block-number> <node> 0 <value> and iB<block-number> 0 <node>

<value>
Each current source is split into two components: from VDD to ideal ground and from ideal ground to VSS.

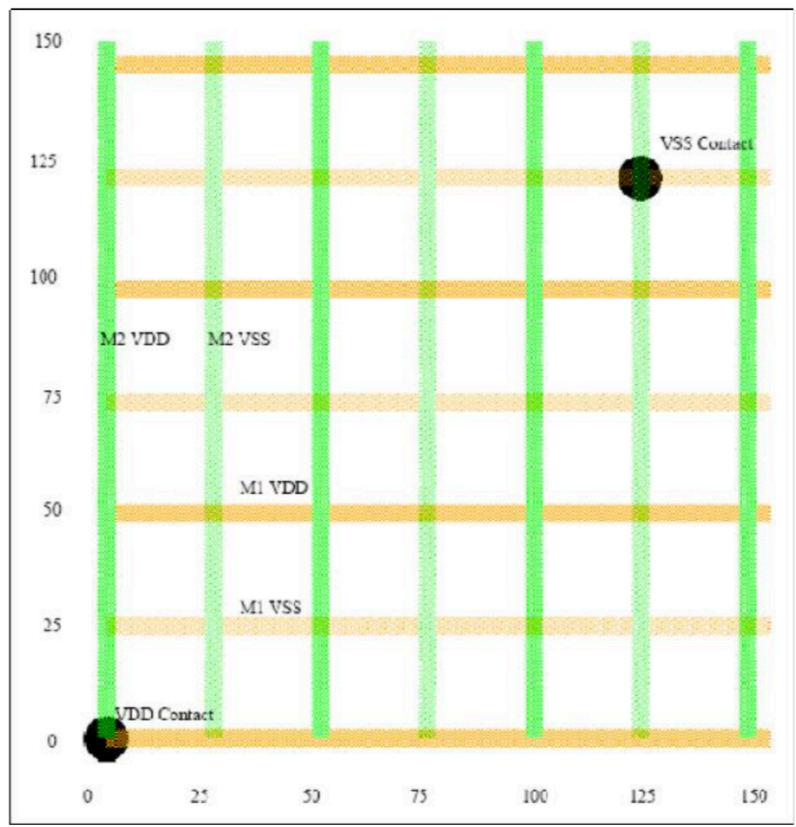

**Fig. 6 Small power grid sample.**

Figure 6 shows a sample small power grid. Reference [10] provides more de- tails of this model.

## 4.2 Effective jL Product Extraction

Blech length effect can be used to quickly filter out a great number of EM- immortal wires which do not require further consideration. The remaining EM- mortal wires will be processed by a more time-consuming EM lifetime calculation procedures. The studies of Blech length effect usually consider simple straight point-to-point wires, for which the values of j and L can be easily determined. For complex interconnect topologies the accumulated jL product along the longest possible path is usually taken. However, in [11], the authors show that it is non- trivial to extract jL product in a complex interconnect topology, and they develop an effective jL extraction strategy based on EM physics, as stated in (3). In (3), $j_k$ and $L_k$ are the current density and length of a wire segment. The sum is taken over all possible paths in the network, with $(jL)_{eff}$ being the maximum of these sums. The effective jL product matches well with experimental results in [11].

$$(jL)_{eff} = \max\left(\sum_k j_k L_k\right)$$

(3)

The effective jL product extraction (3) can be applied to any complex inter- connect structure. But instead of computing all paths in a power grid, the effective jL can be obtained as follows. For any power wire segment, find the longest con- sistent current path that contains it and compute the sum of individual jL products of all wire segments along the path. For example, in Figure 7, paths A, B and C are longest consistent current paths in the given power track, so we have (4).

$$\begin{aligned} jL_{1,eff} &= jL_{2,eff} = jL_{3,eff} = jL_{pathA} = j_1 L_1 + j_2 L_2 + j_3 L_3 \\ jL_{4,eff} &= jL_{5,eff} = jL_{pathB} = j_4 L_4 + j_5 L_5 \\ jL_{6,eff} &= jL_{7,eff} = jL_{pathC} = j_6 L_6 + j_7 L_7 \end{aligned}$$

(4)

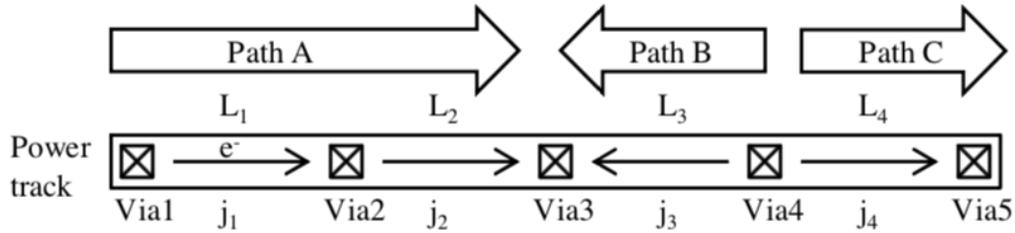

**Fig. 7 Effective jL extraction.**

## 5. EM Lifetime Calculation

### 5.1 Analytical EM lifetime

The Black's equation given by (1) has its limitations, since it ignores some physical factors that can substantially affect EM lifetime such as critical stress for void nucleation, material effective modulus and diffusivity. Many physics-based EM lifetime calculations are developed in literature. We adopt the method pre- sented in [12], which obtains EM lifetime based on the void nucleation and growth model. It also provides EM lifetime calculation for complex interconnect topolo- gies, which can be applied to a power grid.

Here, we briefly explain how to calculate EM lifetime of a power grid wire segment. Detailed discussion can be found in [13]. In Fig. 4.3, arrows denote elec- tron flow in a wire segment. First, each wire segment EM failure is associated with its cathode via. So we have (5).

$$\begin{aligned} t_{L1} &= t_{via1} \\ t_{L2} &= t_{via2} \\ t_{L3} &= t_{L4} = t_{via4} \end{aligned} \tag{5}$$

Next, EM lifetime of a via depends on the metal atomic flux divergences con- tributed by all the connected wire segments. Due to the flux divergence, a void is nucleated when threshold stress is reached. The time for a void nucleation com- prises the first part of (6). After a void is nucleated, it keeps growing until it reaches a critical length such that EM failure occurs. The time for void growth comprises the second part of (6).

$$t_{life} = \left( \frac{\sigma_{void} \Omega}{\rho e Z^*} \sqrt{\frac{\pi}{4}} \sqrt{\frac{kT}{B\Omega}} \frac{\sum_i \sqrt{D_i}}{\sum_i D_i j_i} \right)^2 + \frac{L_c kT}{\rho e Z^*} \frac{1}{\sum_i D_i j_i} \tag{6}$$

In (6), $\sigma_{void}$ is the critical stress required to nucleate a void, $j_i$ and $D_i$ are the current density and the effective diffusivity of each wire segment i connected to a via, B is the effective modulus of the materials surrounding the metal, $L_c$ is the critical void length, and T is the temperature. In reality, if the actual chip tempera- ture profile is available, it should be used for EM lifetime calculation. However, here we assume temperature is uniformly fixed to 300K. The effective diffusivity D is computed using equation (7):

$$D = D_0 \cdot e^{-\frac{E_a}{kT}} \cdot \frac{w}{h} \qquad (7)$$

In (7), w is the wire width and h is its height. When all $D_i$ values are the same, which is typical in a power grid, the lifetime of a wire segment and thus the lifetime of its cathode via is inversely proportional to the current density through the via. The parameter values we used are listed in Table 1.

**Table 1. Physical Parameters**

| . Parameters | Values |
| --- | --- |
| $\sigma_{void}$ | 40MPa |
| $Z^*$ | 1 |
| B | 28GPa |
| $\Omega$ | $1.18 \times 10^{-29} m^{-3}$ |
| $\rho$ | 2.1µΩ·cm (metal 5, 6), 3.9µΩ·cm (metal 2, 3, 4), 4.8µΩ·cm (metal 1) |

Wires are classified into three subgroups based on their analytical lifetime: EM-safe, EM-sensitive and EM-weak. The classification criteria are listed below.

EM-safe: $t_{life} > (1+\beta) \, t_{req}$; EM-weak: $t_{life} < t_{req}$; EM-sensitive: $t_{req} \leq t_{life} \leq (1+\beta) t_{req}$,

where $t_{req}$ is the required lifetime, assumed here to be 10 years. β is the lifetime slack threshold; its value is selected according to the maximum lifetime reduction found by the variation-aware analysis. The wires with lifetime $t_{life} > (1+\beta) \, t_{req}$ are EM-safe and should not fail even if they experience the worst process variation ef- fects. Wires with $t_{req} \leq t_{life} \leq (1+\beta) t_{req}$ are considered EM-sensitive and their status is determined by the variation-aware analysis using the compact model. A user can either use a pre-estimated upper bound of β, or run several iterations of our flow to tune the value of β. For example, if initially the value of β is set to 10%, and later variation-aware analysis reports a 12% maximum lifetime reduction, then the user would adjust β value to a number greater than 12%. In practice, β is within a range of 10%~15%, thus the convergence of tuning β is not considered a problem in our work. All EM-sensitive wires will then be processed by the varia- tion-aware EM analysis.

## 5.2 Global Current Redistribution

CMP/EPE variations not only locally affect EM lifetime, but also cause global power grid current redistribution from full-chip analysis point of view. However, we need to know the exact CMP/EPE variations on the power grid to study those redistribution effects.

For experimental purpose, to create the environment for studying global current distribution changes due to resistance change, we apply Monte-Carlo simulation. Monte-Carlo simulation samples are generated by first randomly throwing some darts on the P/G network. Each such selected resistor, referred to as *c-resistor*, is a variation center. Its own resistance and those of a few wires around it are changed. Each *c-resistor* randomly picks a value according to a Gaussian distribution $N(\mu,\sigma^2)$, where $\mu$ is mean and $\sigma^2$ is variance. This variation then propagates to its neighbors with decreasing magnitude as the distance to the *c-resistor* increases. The assumption is that wires in close layout proximity have similar imperfections due to similar process conditions. Figure 8 gives an example.

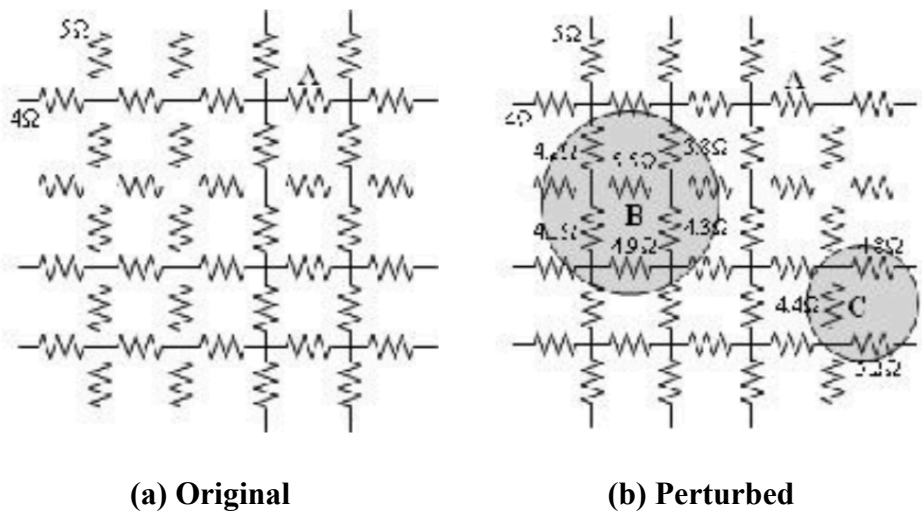

        **(a) Original**                      **(b) Perturbed**

**Fig. 8. P/G network.**

In Figure 8, the nominal value of each horizontal resistor is 5Ω and of vertical resistor is 4Ω. *A* is an EM-sensitive wire. *B* and *C* are *c-resistors* in the perturbed network. The perturbed resistance values are shown in Figure 8 (b). An EM- sensitive wire segment may also become a *c-resistor*.

Currents of the EM-sensitive wires have to be computed in the perturbed grid. Solving Kirchhoff's Current Law (KCL) or Kirchhoff's Voltage Law (KVL) equa- tions for the whole system each time a perturbed network is generated is infeasible due to a huge computational cost. Here, knowing the node voltages in the nominal network, we apply random walk method [14-17] to compute current flows in its per- turbed variant. Random walk is a technique that allows localized computation; its principle is briefly described below using Figure 9.

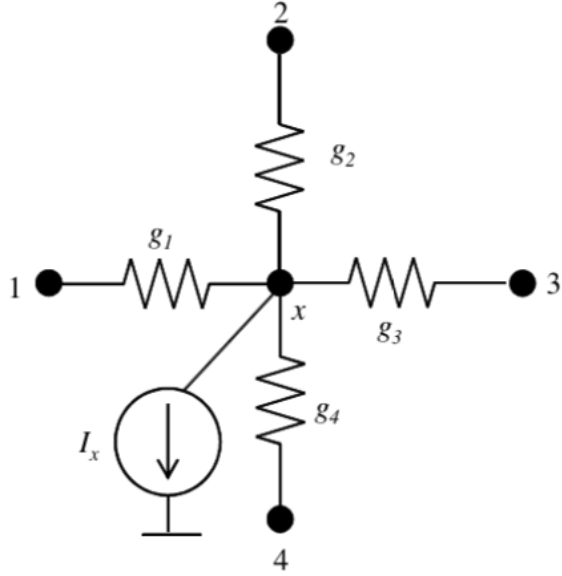

**Fig. 9 Random walk from node x.**

A random walk starts from any node $x$, and then travels to some adjacent node according to a certain probability as in (8) which depends on electrical conduc- tance between the two nodes. In (8), $V_x$ is the voltage of the node $x$, degree($x$) is the number of nodes adjacent to $x$, $g_i$ is the electrical conductance between node $i$ and $x$, $I_x$ is the current load connected to $x$ ($I_x$=0 if no load is connected to $x$). The travelling continues until a boundary node (a ground node or an ideal voltage source node) is reached. Multiple travels are executed from a given node, and then the average value is taken as its node voltage. So the voltages of those nodes of in- terest (nodes of EM-sensitive wires) in a perturbed network can be obtained with- out solving the whole system.

$$V_x = \sum_{i=1}^{\text{degree}(x)} \frac{g_i}{\sum_{j=1}^{\text{degree}(x)} g_j} V_i - \frac{I_x}{\sum_{j=1}^{\text{degree}(x)} g_j}$$

(8)

Lifetimes of all EM-sensitive wires are then recalculated for each perturbed network. Two distributions are obtained: lifetime distribution of each EM- sensitive wire and cumulative distribution function (CDF) of the full-chip EM re- liability.

## 6. Experimental Results

We implemented our tool in C++ and tested it with several publicly released IBM P/G benchmarks [18-22]. Experiments were carried on GUN/LINUX server with Intel Xeon E5440

2.83GHz CPU and 16GB memory. The specification required EM lifetime (referred to as $t_{req}$) was set to 10 years and the temperature was assumed to be 300K for all experiments.

In Figure 10, we compare $jL$ filter results from the accumulated and effective $jL$ product extraction for metal 3 layer in IBM PG1 benchmark. An obvious reduction in EM-mortal wires can be observed in effective $jL$ product distribution. This is because the accumulated $jL$ extraction uses the longest possible path for an entire power track, thus a large number of wires are unnecessarily classified as EM-mortal. This imposes an onerous burden on further EM analysis and significantly diminishes the efficiency of fast $jL$ filter.

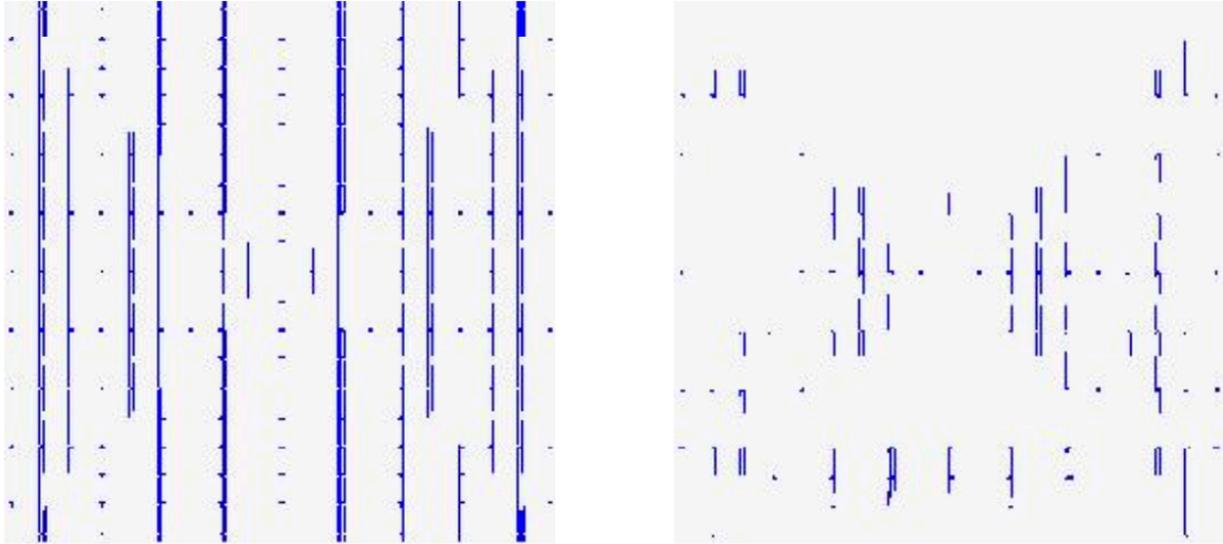

(a) Accumulated mortal map          (b) Effective mortal map

**Fig. 10 EM-mortal map for IBM PG1.**

Moreover, there is a risk of missing EM-mortal wires using the accumulated $jL$ product. For example, in Figure 11, two currents with the same magnitude but opposite directions are flowing from two ends of a power track and meet at the middle point. The accumulated $jL$ extraction will classify this wire as EM-immortal, because the accumulated $jL$ product would be zero, whereas the effective $jL$ product extraction suggests computing $jL$s for the two consistent paths separately. The quantitative results of $jL$ filter are listed in Table 2.

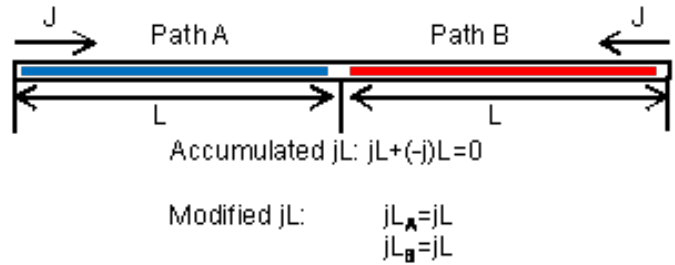

**Fig. 11 $jL$ extraction comparison.**

When considering global effects of process variations, we assume that 10% of wires experience variations, with each *c-resistor* $R_i$ following $N(R_i, 0.001R_i^2)$ distribution. We simulated the global effects of process variation on a randomly selected EM-sensitive wire, named $W_2$, taken from IBM PG2 benchmark.

All these results are quantitatively summarized in Table 2. In Table 2, the sub column labeled *A* under column *Mortal* provides the accumulated *jL* extraction values, the sub column labeled *E* refers to effective *jL* extraction and the sub column labeled *Miss* gives the number of EM-mortal wires missed by the accumulated *jL* extraction. The column labeled *Avg. prob. no fail* is the average probability that EM-sensitive wires will not fail.

**Table 2 EXPERIMENTAL RESULTS.**

| Benchmarks | Total #wires | Mortal | | | Worst MTTF (year) | #EM-sensitive | Avg. prob. No failure | Avg. CMP tolerance | Avg. EPE tolerance |
| --- | --- | --- | --- | --- | --- | --- | --- | --- | --- |
| | | A | E | Miss | | | | | |
| IBM PG1 | 30027 | 25196 | 10032 | 623 | 4.07 | 46 | 32.7% | 15.74% | 8.50% |
| IBM PG2 | 208325 | 111245 | 23124 | 77 | 4.94 | 63 | 89.7% | 16.24% | 15.41% |
| IBM PG3 | 1401572 | 33872 | 33309 | 546 | 4.90 | 917 | 82.0% | 14.37% | 12.04% |
| IBM PG4 | 1560645 | 198103 | 66226 | 0 | 5.53 | 554 | 74.6% | 12.53% | 11.06% |

## 7. Conclusions

In this paper, we show significant impact on electromigration reliability from uneven current distribution and multiple variation effects. The uneven current distribution is modeled using FEM approach and compact modeling is provided. For chip-level variation effects, we propose a Gaussian variation distribution that simulates the variation effects, and random walk algorithm is used to efficiently estimate the chip-level impacts. Experimental results show substantial EM degradation when variation effects are present.

## References

[1] M. Lin, N. Jou, W. Liang and K. C. Su, "Effect of Multiple Via Layout on Electromigration Performance and Current Density Distribution in Copper Interconnect," *International Reliability Physics Symposium*, pp. 844-847, 2009.

[2] N. Raghavan and C. M. Tan, "Statistical Modeling of Via Redundancy Effects on Interconnect Reliability," *International Symposium on the Physical and Failure Analysis of Integrated Circuits*, pp. 1-5, 2008.

[3] D. A. Li, M. Marek-Sadowska and S. R. Nassif, "T-VEMA: A Temperature- and Variation-Aware Electromigration Power Grid Analysis Tool," in IEEE Transactions on Very Large Scale Integration (VLSI) Systems, vol. 23, no. 10, pp. 2327-2331, Oct. 2015.

[4] http://en.wikipedia.org/wiki/Activation_energy.

[5] R. Reis. Circuit Design for Reliability, Springer, New York, 2015.

[6] R. L. D. Orio, "Electromigration Modeling and Simulation," doctoral dissertation, Institute for Microelectronics, TU Wien, 2010.

[7] D. Li, M. Marek-Sadowska and S. R. Nassif, "A Method for Improving Power Grid Resilience to Electromigration-Caused Via Failures," to appear in *IEEE Trans. on VLSI Systems*.

[8] D. Dalleau, "3-D Time-depending Simulation of Void Formation in Metallization Structures", Doctoral thesis, University of Hannover, 2003.

[9] W. Li, C. M. Tan and N. Raghavan, "Dynamic simulation of void nucleation during electromigration in narrow integrated circuit interconnects," *Journal of Applied Physics* 105, 014305 (2009).

[10] D. Li, et al. Variation-aware electromigration analysis of power/ground networks. Proceedings of the International Conference on Computer-Aided Design, 2011.

[11] M. Lin, N. Jou, W. Liang and K. C. Su. Effect of Multiple Via Layout on Electromigration Performance and Current Density Distribution in Copper Interconnect. *IEEE 47th Annual International Reliability Physics Symposium*, Montreal, 2009, pp. 844-847.

[12] B. Li, J. Gill, C. J. Christiansen, et. al. Impact of Via-Line Contact on Cu Interconnect Electromigration Performance. *IEEE 43rd Annual International Reliability Physics Symposium*, San Jose, 2005, pp. 24-30.

[13] D. Li and M. Marek-Sadowska, "Estimating true worst currents for power grid electromigration analysis," Fifteenth International Symposium on Quality Electronic Design, Santa Clara, CA, 2014, pp. 708-714.

[14] F. M. Serry, D. Dawson, "Minimizing Dishing and Erosion in Copper CMP," http://www.veeco.com/pdfs/database_pdfs/minimizing_de_in_copper_cmp_45.pdf.

[15] F. Chen et al., "Investigation of emerging middle-of-line poly gate-to-diffusion contact reliability issues," 2012 IEEE International Reliability Physics Symposium (IRPS), Anaheim, CA, 2012, pp. 6A.4.1-6A.4.9.

[16] D. Li, et al. On-chip em-sensitive interconnect structures. Proceedings of the international workshop on system level interconnect prediction, 2010.

[17] S. R. Nassif, "Power grid analysis benchmarks," *Asia and South Pacific Design Automation Conference*, 2008, pp. 376-381.

[18] Q. Wang, et al. Research on phase locked loop based on DSP and CPLD. Mechanical & Electrical Engineering Magazine, 2007.

[19] D. Li, et al. "Layout Aware Electromigration Analysis of Power/Ground Networks". Circuit Design for Reliability, 2015.

[20] C. W. Chang, Z. –S. Choi, C. V. Thompson, et.al, "Electromiration resistance in a short three-contact interconnect tree," *Journal of Applied Physics* 99, 094505 (2006).

[21] H. Qian, S. R. Nassif, and S. S. Sapatnekar, "Power Grid Analysis Using Random Walks," *IEEE Trans. on CAD*, vol. 24, no. 8, August 2005, pp. 1204-1224.

[22] S. M. Alam, C. L. Gan, C. V. Thompson, et al, "Reliability computer-aided design tool for full-chip electromigration analysis and comparison with different interconnect metallizations," *Microelectronics Journal* 38 (2007) 463-473.